# Reduced Palm Intensity for Track Extraction

Ali Önder Bozdoğan, *Member, IEEE*, Roy Streit, *Fellow, IEEE*, and Murat Efe, *Member, IEEE*

*Abstract*—The pair correlation function is introduced to target tracking filters that use a finite point process target model as a means to investigate interactions in the Bayes posterior target process. It is shown that the Bayes posterior target point process of the probability hypothesis density (PHD) filter—before using the Poisson point process approximation to close the recursion—is a spatially correlated process with weakly repulsive pair interactions. The reduced Palm target point process is introduced to define the conditional target point process given the state of one or more known targets. Using the intensity function of the reduced Palm process, an approximate two-stage pseudo maximum a posteriori track extractor is developed. The proposed track extractor is formulated for the PHD filter and implemented in a numerical study that involves tracking two close-by targets. Numerical results demonstrate improvement in the mean optimal subpattern assignment statistic for the proposed track extractor in comparison to the Gaussian mixture PHD filter's state extractor.

*Index Terms*—Reduced Palm intensity, Palm filter, Conditional intensity, Pair correlation, Intensity filter, PHD filter, Track extraction.

## I. Introduction

A set of target tracking filters based on finite point process models such as the probability hypothesis density (PHD) filter [1], the intensity filter (iFilter) [2], and the cardinalized PHD (CPHD) filter [1] bypass the need to explicitly enumerate measurement-to-target assignments by approximating the Bayes posterior target process with a finite point process model in which all targets share a common probability density function (pdf) which is proportional to the intensity function of the posterior target process. This approximation ignores correlations between pairs of target states and any higher order interactions that involve groups of targets. The present paper is concerned with pair correlations that exist between targets in the Bayes posterior point process of the PHD filter before it is approximated with a Poisson point process (PPP) in order to close the recursion that defines the filter algorithm. Explicit expressions for the pair correlation function of the Bayes posterior point process are derived from the generating functional of the joint target-measurement point process.

The existence of pair, or spatial[1], correlation in the Bayes posterior target process implies that track extraction methods should somehow compensate for the correlation between detected targets. Although of little practical significance in many situations, ignoring this correlation will lead to biased state estimates, especially for closely spaced targets. To compensate for the spatial correlation, [3] introduced methods leading to a finite point process called the reduced Palm process [4], [5, Chapter 13], [6]. The intensity function of the reduced Palm process acts like a whitening filter, meaning that the reduced Palm process uses conditioning on detected targets to reduce the intensity of the Bayes posterior intensity function. Simply said, the reduced Palm process is a kind of subtractor, or notched intensity filter. While the intensity function collapses the multitarget state into a single target state space, the reduced Palm process separates individual detected targets from the other targets.

---

[1] In this paper, the term spatial refers to the full target state which may involve velocity components.

Derivation of the conditional intensity function of the reduced Palm process requires the calculation of second or higher factorial moments of the Bayes posterior process. For the PHD filter, it is seen in [3] that the information encoded in the second factorial moment is of value for resolving a target pair that may be otherwise merged in the intensity, or PHD, function. While earlier work in [7] also studies the second-order statistics of multitarget systems in order to devise a second order multitarget tracking filter which propagates not only the intensity function but also the covariance density of the random track set, its conclusions suggest that computationally tractable second order multitarget filters are unlikely to be devised. Therefore, [7] offers an alternative path to generalize Kalman filtering for multitarget systems. This work is further developed in [8] and [9]. This line of work eventually leads to the partially higher order (i.e., "first order in the states of individual targets, but of higher order in the target number" [10]) target process model for the CPHD filter and its approximations such as [11]. In contrast, [3] does not specifically aim to derive a new filter based on an enhanced target process model. It exploits the information encoded in the second or higher factorial moments of the Bayes posterior target process for the purpose of track extraction. Its results are general and applicable to cardinal filters as well.

Building on the results of [3], this paper provides three main contributions. Firstly, in Section V, using the second factorial moment of the Bayes posterior process, the exact form of the pair correlation function of the PHD filter is derived. As seen in Section V.B, this proves analytically that the Bayes posterior process of the PHD filter before the PPP approximation is inherently a spatially correlated process. This is because of the fact that the pair correlation function of the Bayes posterior process is less than or equal to one for any pair of target states. In contrast, lack of correlation in the usual statistical sense corresponds to the case where the pair correlation function is strictly equal to one [4]. A finite point process that has a pair correlation function which is less than one for any arbitrary pair of points will be called a weakly repulsive point process (see, e.g., [12, Section 6.1.1] for the definition of repulsive and attractive point processes).

Secondly, in Section VI, a two-stage pseudo maximum a posteriori (pseudo-MAP) track extractor for a general finite point process multitarget model is derived. The first stage of the track extractor estimates the canonical, or cardinal, number of the target process, and the second stage sequentially extracts marginal target pdfs due to extracted targets using a general bivariate target-measurement process probability generating functional (PGFL). While the two-stage track extraction process is well-known (see, e.g., [1]), the introduction of the conditional intensity function for the purpose of sequential peak determination and the introduction of the reduced Palm process for the purpose of individual target pdf extraction are novel contributions of this work. The track extractor for the PHD filter is formulated.

Thirdly, in Section VII, the paper derives the reduced Palm intensity function of the Bayes posterior process of the PHD filter and uses it to carry out the multi-dimensional optimization required to find the pseudo-MAP estimate of the extracted target states. This sequential procedure uses a series of single target optimizations. The introduction of a modulation term which is called herein the Palm corrector is the last main contribution of this paper. The Palm corrector is a subtractive tool that acts similarly to a spatial whitening filter (a name adopted from the traditional time series signal processing literature). It is a general tool which can also be applied to other tracking filters.

The Palm track extractor does not directly depend on explicit [13-18] or implicit [19],[20] clustering methods. The Palm method proposed in this paper is similar to the CLEAN algorithm of radio astronomy [21], which removes targets from the intensity function sequentially using the shape of the intensity surface around expected target states [22],[23].

The outline of this paper is as follows: The PGFL of a finite point process and its functional derivatives are described in Section II. The PGFL of the reduced Palm process is derived in Section III. In Section IV, the joint PGFL of two finite point processes is given and used to find the PGFL of the Bayes posterior of one point process conditioned on realizations of another one. Additionally, the pair correlation function is defined in Section IV. In Section V, the pair correlation function for the PHD filter is derived. The two-stage track extraction algorithm is presented in Section VI, and the track extraction algorithm is formulated for the PHD filter in Section VII. In Section VIII, numerical examples are presented, and concluding remarks are given in Section IX.

## II. PROBABILITY GENERATING FUNCTIONAL

The event space $\mathcal{E}(S)$ of a simple finite point process $\Xi$ is the set of all ordered pairs of the form $(n,\{s_1,\ldots,s_n\})$. The points $s_i$ are in the state space $S$, which is typically taken to be a specified subset of $\mathbb{R}^{\dim(S)}$. The event space is called the "Grand Canonical Ensemble," and the number $n$ is called the canonical, or cardinal, number. For $n=0$, the event is $(0,\varnothing)$. For $n \geq 1$, the event corresponds to $n!$ equally likely ordered events of the form $(n, s_{\sigma(1)}, \ldots, s_{\sigma(n)})$, $\sigma \in Sym(n)$, where $Sym(n)$ denotes the set of all permutations of the first $n$ positive integers.

Background discussions for finite point processes can be found in the authoritative text [5]. Consider a real-valued function $h$ on the state space $S = \mathbb{R}^{\dim(S)}$, the probability generating functional (PGFL) of $\Xi$ is defined as:

$$G[h] = E[h] = \sum_{n=0}^{\infty} \int_{S^n} \left( \prod_{i=1}^{n} h(s_i) \right) p(n, s_1, \ldots, s_n) ds_1 \ldots ds_n \tag{1}$$

Simply put, the PGFL is the expectation of the random product $\prod_{i=1}^{n} h(s_i)$, and it is only evaluated for functions $h$ such that (1) is absolutely convergent. It is shown in [24] (Theorem 4.1) that for symmetric probability distributions, i.e. distributions that are invariant under point order permutations, the PGFL determines a finite point process uniquely. The next section describes a method of extracting information from the PGFL by functional differentiation. Functional differentiation is essentially the same as in the classical Calculus of Variations, but with an additional second limiting step involving a test sequence for a Dirac delta (discussions of test sequences are widely available, e.g., see [25]).

### A. Derivatives of the Probability Generating Functional

The functional derivative of $G[h]$ with respect to the variation $w$ is defined by

$$\frac{\partial G}{\partial w}[h] = \lim_{\varepsilon \to 0} \frac{d}{d\varepsilon} G[h + \varepsilon w] \tag{2}$$

The variation $w$ is a specified bounded real-valued function on $S$ and it is assumed that $\sup_{s \in S} |h(s)| < 1$. Here, we assume that $w$ is a function in a test sequence for the Dirac delta (see, e.g., [25]). Using (1) and (2) gives

$$G[h + \varepsilon w] = \sum_{n=0}^{\infty} \int_{S^n} \prod_{i=1}^{n} \left[ h(s_i) + \varepsilon w(s_i) \right] p(n, s_1, \ldots, s_n) ds_1 \ldots ds_n \tag{3}$$

Since integrals and sums are absolutely convergent, taking the limit in (2) gives

$$\frac{\partial G}{\partial w}[h] = \sum_{n=1}^{\infty} \sum_{k=1}^{n} \int_{S^n} w(s_k) \left( \prod_{i=1, i \neq k}^{n} h(s_i) \right) p(n, s_1, \ldots, s_n) ds_1 \ldots ds_n \tag{4}$$

Now, the limit of (4) taken over the test function sequence $w$ for the Dirac delta $\delta_x(s) \equiv \delta(s-x)$ with a point mass at $s = x \in S$ defines the derivative of the PGFL as

$$\frac{\partial G}{\partial x}[h] = \frac{\partial G}{\partial \delta_x}[h] = \frac{\partial G}{\partial w}[h]\bigg|_{w(.)=\delta_x(.)} = \sum_{n=1}^{\infty}\sum_{k=1}^{n}\int_{S^n} \delta_x(s_k)\left(\prod_{i=1,i\neq k}^{n} h(s_i)\right)p(n,s_1,\ldots,s_n)ds_1\ldots ds_n \quad (5)$$

Using the sampling property of the Dirac delta, the symmetric structure of $p(n,s_1,\ldots,s_n)$ and relabeling arguments give:

$$\frac{\partial G}{\partial x}[h] = \sum_{n=1}^{\infty} n \int_{S^{(n-1)}} \left(\prod_{i=2}^{n} h(s_i)\right) p(n,x,s_2,\ldots,s_n) ds_2 \ldots ds_n \quad (6)$$

Higher derivatives can be derived using the same mechanics. The result is [26]

$$\frac{\partial^k G}{\partial x_1 \ldots \partial x_k}[h] = \sum_{n=k}^{\infty} n(n-1)\ldots(n-k+1) \int_{S^{(n-k)}} \left(\prod_{i=k+1}^{n} h(s_i)\right) p(n,x_1,\ldots,x_k,s_{k+1},\ldots,s_n) ds_{k+1} \ldots ds_n \quad (7)$$

Derivatives of the PGFL with respect to any finite number of Dirac deltas provide a mechanism to recover the pdf of the point process. For example, the $k$th derivative of the PGFL evaluated at $h = 0$ gives:

$$\frac{\partial^k G}{\partial x_1 \ldots \partial x_k}[0] = k!\, p(k,x_1,\ldots,x_k) \quad (8)$$

This is the pdf of the unordered event $\{x_1,\ldots,x_k\}$.

The PGFL evaluated at $h(.) = x$ generates the probability generating function (PGF) of the canonical number of points of the point process [26]:

$$F(x) = G[h]\big|_{h(.)=x} = \sum_{n=0}^{\infty} x^n p(n) \quad (9)$$

The expected number of points is the derivative of $F(x)$ evaluated at $x = 1$, that is, $E[N] = F'(1)$.

*B. Factorial Moments of the PGFL*

The special case of (6) as $h(.)$ approaches 1 is

$$m_{[1]}(x) = \lim_{h\to 1} \frac{\partial G}{\partial x}[h] = \sum_{n=1}^{\infty} n \int_{S^{(n-1)}} p(n,x,s_2,\ldots,s_n) ds_2 \ldots ds_n \quad (10)$$

The function $m_{[1]}(x)$ is called the first moment of the point process. In the larger literature, it is also called as the intensity function. In the tracking literature [1], it is often referred to as the PHD function. It gives the expected number of points in the point process per unit state space $S$.

With Dirac deltas at distinct locations, the higher derivatives evaluated with $h(.) \to 1$ give factorial moments of the point process:

$$m_{[k]}(x_1,\ldots,x_k) = \lim_{h\to 1} \frac{\partial^k G}{\partial x_1 \ldots \partial x_k}[h] = \sum_{n=k}^{\infty} n(n-1)\ldots(n-k+1) \int_{S^{(n-k)}} p(n,x_1,\ldots,x_k,s_{k+1},\ldots,s_n) ds_{k+1} \ldots ds_n \quad (11)$$

Factorial moments [24] are the key to understanding the Palm process and pair (and higher order) correlation functions.

In what follows, the limit $h(.) \to 1$ will be dropped for notational convenience, i.e. $\frac{\partial^k G}{\partial x_1 \ldots \partial x_k}[1] \equiv \lim_{h\to 1} \frac{\partial^k G}{\partial x_1 \ldots \partial x_k}[h]$.

### III. THE REDUCED PALM PROCESS

The Palm distribution formalizes conditioning on a point of the process. Intuitively speaking, it describes the distribution of the population given that there is a point at a specified location. The Palm process includes this point in realizations of the process. The theory was first developed by C. Palm in 1943 in early work in the area of telephone communication (see [27] for an English translation of Palm's original paper). It was further developed and generalized to several dimensions by others (see [5, Chapter 13], [28], and references therein). For the purpose of track extraction however, what is of interest is the distribution of the population *excluding* the point given at the specific location. This distribution defines the "reduced" Palm process.

Let the pdf $p(n, v_1, \ldots, v_n)$ be described as:

$$p(n, v_1, \ldots, v_n) = \lim_{dV \to 0} \frac{\Pr \left\{ \begin{array}{c} \text{there are exactly } n \text{ points and } i^{th} \text{ point} \\ \text{is in the ball of radius } \varepsilon \text{ centered at } v_i \\ \text{of volume } dV \text{ for } i = 1, \ldots, n \end{array} \right\}}{(dV)^n} \quad (12)$$

Because of the symmetric structure of $p(.)$, given a fixed set of points $x^k = \{x_1, \ldots, x_k\}$, $k \geq 1$, and a set $v^t = \{v_1, \ldots, v_t\}$, $t \geq 0$, the probability that an event produced by the process has $t + k$ distinct points in infinitesimal balls centered at the points of the set union $X = x^k \cup v^t = \{x_1, \ldots, x_k, v_1, \ldots, v_t\}$ is $p(t+k, v_1, \ldots, v_t, x_1, \ldots, x_k) \binom{t+k}{k} k! \, dV^{n+k}$.

In this expression, the coefficient $\binom{t+k}{k} k!$ counts the number of ways in which $t + k$ members can be assigned to $k$ fixed balls centered at the points of $x^k$. Finding the conditional probability of the event that point process has $t$ additional members in infinitesimal balls centered at the elements of $v^t = \{v_1, \ldots, v_t\}$ given that the event has $k$ distinct members in infinitesimal balls centered at the points of $x^k$ involves integrating out $v_1, \ldots, v_t$ over all possible states and for all canonical numbers $t$. That is, the reduced Palm pdf is defined as:

$$\lim_{\varepsilon \to 0} p(t, v_1, \ldots, v_t \mid \{x_1, \ldots, x_k\}) = \frac{p(t+k, v_1, \ldots, v_t, x_1, \ldots, x_k) \binom{t+k}{k} k!}{\sum_{i=0}^{\infty} \int_{S^i} p(i+k, s_1, \ldots, s_i, x_1, \ldots, x_k) \binom{i+k}{k} k! \, ds_1 \ldots ds_i} \quad (13)$$

provided the limit exists. Let $n = t + k$, then (13) can be rewritten as:

$$\lim_{\varepsilon \to 0} p(t, v_1, \ldots, v_t \mid \{x_1, \ldots, x_k\}) = \frac{p(n, v_1, \ldots, v_t, x_1, \ldots, x_k) n(n-1) \ldots (n-k+1)}{\sum_{i=k}^{\infty} \int_{S^{i-k}} p(i+k, s_1, \ldots, s_i, x_1, \ldots, x_k) i(i-1) \ldots (i-k+1) \, ds_1 \ldots ds_i} \quad (14)$$

The denominator of (13) is equal to the *k*th factorial moment given in (11).

*A. The PGFL of the Reduced Palm Process*

The conditional probability model of the reduced Palm process is defined in (14). With the required index changes, the PGFL of the reduced Palm process is equal to:

$$G[h \mid \{x_1, \ldots, x_k\}] = E[h \mid \{x_1, \ldots, x_k\}] = \sum_{n=k}^{\infty} \int_{S^{n-k}} \left( \prod_{i=1}^{n-k} h(s_i) \right) \frac{p(n, s_1, \ldots, s_t, x_1, \ldots, x_k)}{m_{[k]}(x_1, \ldots, x_k)} n(n-1) \ldots (n-k+1) \, ds_1 \ldots ds_{n-k} \quad (15)$$

Since $m_{[k]}(x_1, \ldots, x_k)$ is independent of the summation and the integration, (15) is proportional to the functional derivative of $G[h]$ with respect to $k$ distinct Dirac deltas located at $x_1, \ldots, x_k$ given by (7). The PGFL of the reduced Palm process is seen to be the ratio of the *k*th derivative of the PGFL with the *k*th factorial moment:

$$G[h|\{x_1,\ldots,x_k\}] = \frac{\frac{\partial^k G}{\partial x_1 \ldots \partial x_k}[h]}{m_{[k]}(x_1,\ldots,x_k)} = \frac{\frac{\partial^k G}{\partial x_1 \ldots \partial x_k}[h]}{\frac{\partial^k G}{\partial x_1 \ldots \partial x_k}[1]} \quad (16)$$

The factorial moment in the denominator acts as a (Bayesian) normalizing factor. In words, (16) says that the normalized functional derivatives of a finite point process are the PGFLs of the reduced Palm processes [5, Proposition 13.2.VI].

Realizations of the reduced Palm process events do not include points of the conditioning set. The Palm process events are formed from the reduced Palm process events by concatenating the conditioning set to each event. The reduced Palm process is more suitable to track extraction purposes than the Palm process itself.

If it exists, the intensity function of the reduced Palm process, at $x_{k+1}$ is given by:

$$m_{[1]}(x_{k+1}|\{x_1,\ldots,x_k\}) = \frac{\frac{\partial^{k+1} G}{\partial x_1 \ldots \partial x_{k+1}}[1]}{\frac{\partial^k G}{\partial x_1 \ldots \partial x_k}[1]} \quad (17)$$

*Example I: The reduced Palm Process for the PPP*

The PGFL of the PPP with the intensity function $f(s)$ is given by [26]:

$$G[h] = \exp\left(\int_S f(s)(h(s)-1)ds\right) \quad (18)$$

Using the results of section 2.A, the functional derivative of $G[h]$ with respect to a Dirac delta at $x$ is

$$\frac{\partial G}{\partial x}[h] = G[h]f(x) \quad (19)$$

Using (15)-(19), it is straightforward to verify the well-known result that, the PGFL of the reduced Palm process is equal to that of the original PPP:

$$\frac{\frac{\partial G}{\partial x}[h]}{\frac{\partial G}{\partial x}[1]} = G[h] \quad (20)$$

Furthermore, the derivatives of the PGFL have the factorized form

$$\frac{\partial^n G}{\partial x_1 \ldots \partial x_n}[h] = G[h]\prod_{i=1}^{n} f(x_i) \quad (21)$$

From (21), it follows that the PGFL of the reduced Palm process conditioned on *k* distinct points is identical to the PGFL of the original PPP:

$$\frac{\frac{\partial^k G}{\partial x_1 \ldots \partial x_k}[h]}{\frac{\partial^k G}{\partial x_1 \ldots \partial x_k}[1]} = G[h] \quad (22)$$

This is a restatement of the well-known independent scattering property of PPPs [2]. It does not hold for general processes.

Another result of the independent scattering property of PPPs is the factorization of its moments. This will lead to the conditional intensity function at an arbitrary point $x_{k+1} \notin \{x_1,\ldots,x_k\}$ to be:

$$m_{[1]}\left(x_{k+1} \mid \{x_1,\ldots,x_k\}\right) = \frac{m_{[k+1]}(x_1\ldots x_{k+1})}{m_{[k]}(x_1\ldots x_k)} = \frac{\prod_{i=1}^{k+1} m_{[1]}(x_i)}{\prod_{i=1}^{k} m_{[1]}(x_i)} = m_{[1]}(x_{k+1}) \quad (23)$$

Therefore, for PPPs, the intensity function at an arbitrary point is independent of its value at any other point that is not in the set $\{x_1,\ldots,x_k\}$. For target tracking purposes, a PPP multitarget model indicates that there is no dependency between distinct target states.

*Example II: The PGFL of the Reduced Palm Distribution for the Independently and Identically Distributed Clusters*

Independently and identically distributed (IID) clusters are the finite point process model of the measurement and the target point processes for the cardinalized PHD filter [10]. Dividing $p(n, s_1,\ldots, s_n)$ into two functions, namely the distribution of the canonical, or cardinal, number with the point mass function $p_N(n)$ and the spatial distribution of each point with the pdf $p(s)$, the PGFL of IID clusters is given by

$$G[h] = \sum_{n=0}^{\infty} p_N(n)\left(\int_S h(s)p(s)ds\right)^n = G^N\left(\int_S h(s)p(s)ds\right) \quad (24)$$

where $G^N(\cdot)$ is the ordinary PGF of the canonical number, $N$, of the process. Let $G'^N(\cdot)$ denote its ordinary derivative. Following Section 2A, the PGFL of the reduced Palm distribution for the IID clusters conditioned on the existence of a point at $x_1$ with $p(x_1) \neq 0$ is

$$G[h \mid \{x_1\}] \equiv \frac{\frac{\partial G}{\partial x_1}[h]}{\frac{\partial G}{\partial x_1}[1]} = \frac{p(x_1)\sum_{n=1}^{\infty} p_N(n)n\left(\int_S h(s)p(s)ds\right)^{n-1}}{p(x_1)\sum_{n=1}^{\infty} p_N(n)n} \quad (25)$$

The term $p(x_1) \neq 0$ can be cancelled. The sum in the numerator is the ordinary derivative of the PGF of the canonical number evaluated at $\int_S h(s)p(s)ds$, while the sum in the denominator is the derivative of the PGF evaluated at one. Thus,

$$G[h \mid \{x_1\}] = \frac{G'^N\left(\int_S h(s)p(s)ds\right)}{G'^N(1)} \quad (26)$$

This gives the conditional intensity function at $x_2$ as

$$m_{[1]}(x_2 \mid x_1) \equiv \frac{\partial}{\partial x_2} G[1 \mid \{x_1\}] = \frac{\frac{\partial}{\partial x_2} G'^N\left(\int_S h(s)p(s)ds\right)\Big|_{h=1}}{G'^N[1]} = \frac{p(x_2)G''^N[1]}{G'^N[1]} = \frac{p(x_2)\sum_{n=2}^{\infty} p_N(n)n(n-1)}{\sum_{n=1}^{\infty} p_N(n)n} \quad (27)$$

where $G''^N(\cdot)$ is the second derivative of $G^N(\cdot)$. The ratio of conditional intensity and the intensity functions at $x_2$ with $p(x_2) \neq 0$ is then given by

$$\frac{m_{[1]}(x_2 \mid x_1)}{m_{[1]}(x_2)} = \frac{\sum_{n=2}^{\infty} p_N(n)n(n-1)}{\left(\sum_{n=1}^{\infty} p_N(n)n\right)^2} \quad (28)$$

For the Poisson distributed canonical number probability mass function (pmf), it is easy to show that the ratio of the conditional intensity and the intensity functions is equal to one. For such case, IID clusters are a PPP. For an arbitrary canonical number pmf however, the ratio can be larger or smaller than one. A simple example to demonstrate this is the uniform canonical number pmf. Assuming a uniform distribution on the number of points from zero to $K \geq 1$, the value of the ratio (28) can be seen to be greater than one for $K > 4$, less than one for $K < 4$ and equal to one for $K = 4$. For general IID clusters, therefore, the factorial moments are not the product of first moments. Moreover, (28) shows that the interaction in the factorial moments at an arbitrary pair of points $x_1$ and $x_2$ (with $p(x_1) \neq 0$ and $p(x_2) \neq 0$) does not depend on the separation between the two points, that is, the interaction between points is global, not local. In target tracking applications, a direct result of this fact is the "spooky" interaction between widely separated targets reported in [30].

## IV. THE BAYES POSTERIOR PGFL FOR TWO PROCESSES

Let $Y$ be a finite point process with events $(m, y_1, \ldots, y_m) \epsilon \mathcal{E}(Y)$, where the measurement space $Y$ is in general unrelated to the target state space $S$. Extending the definition (1) of the PGFL for $\Xi$ to the joint point process $(Y, \Xi)$ with events in the Cartesian product space $\mathcal{E}(Y) \times \mathcal{E}(S)$ gives

$$G^{Y\Xi}[g,h] = \sum_{n=0}^{\infty}\sum_{m=0}^{\infty} \int_{Y^m}\int_{S^n} \left(\prod_{i=1}^{n} h(s_i)\right)\left(\prod_{j=1}^{m} g(y_j)\right) p(m, y_1, \ldots, y_m, n, s_1, \ldots, s_n) dy_1 \ldots dy_m ds_1 \ldots ds_n \qquad (29)$$

The products in (29) are defined to be one for $m = 0$ and $n = 0$. For $h(.) = 1$ or $g(.) = 1$, the PGFL of the joint process reduces to the PGFL of the single process.

$$G^{Y}[g] = G^{Y\Xi}[g,1] \text{ and } G^{\Xi}[h] = G^{Y\Xi}[1,h] \qquad (30)$$

Taking the functional derivatives of $G^{Y\Xi}[g,h]$ with respect to Dirac deltas at $z_1, \ldots, z_k$ in $Y$ and taking $g(.) = 0$ gives

$$\frac{\partial^k G^{Y\Xi}}{\partial z_1 \ldots \partial z_k}[0,h] = k! \sum_{n=0}^{\infty} \int_{S^n} \left(\prod_{i=1}^{n} h(s_i)\right) p(k, z_1, \ldots, z_k, n, s_1, \ldots, s_n) ds_1 \ldots ds_n \qquad (31)$$

Similarly, using (29),

$$\frac{\partial^k G^{Y\Xi}}{\partial z_1 \ldots \partial z_k}[0,1] = k! p^{Y}(k, z_1, \ldots, z_k) \qquad (32)$$

The pdf of the conditional process $\Xi | Y$ is defined by

$$p^{\Xi|Y}(\xi | \nu) \equiv \frac{p^{Y\Xi}(\nu, \xi)}{p^{Y}(\nu)} \qquad (33)$$

where $\nu$ is a realization of $Y$. Therefore, similarly to the PGFL of the reduced Palm process, the PGFL for the Bayes posterior process is the ratio of (31) and (32):

$$G^{\Xi|Y}[h | \nu] = \frac{\dfrac{\partial^k G^{Y\Xi}}{\partial z_1 \ldots \partial z_k}[0,h]}{\dfrac{\partial^k G^{Y\Xi}}{\partial z_1 \ldots \partial z_k}[0,1]} \qquad (34)$$

The functional derivatives of the Bayes posterior PGFL with respect to Dirac deltas with $h(.) \to 1$ give its factorial moments. The first factorial moment is the intensity function, or the PHD function, and gives the expected target count per unit space. From (30) and (31), the intensity function is

$$m_{[1]}(x|\nu) = \frac{\sum_{n=1}^{\infty} n \int_{S^{n-1}} p^{\Xi|Y}(k,z_1,\ldots,z_k,n,x,s_2,\ldots,s_n)ds_2\ldots ds_n}{p^Y(k,z_1,\ldots,z_k)}. \tag{35}$$

*A. The reduced Palm intensity function of the Bayes posterior process*

From (17), the reduced Palm intensity function of the Bayes posterior process at $x_{N+1} \notin \{x_1,\ldots,x_N\}$ conditioning on the existence of the set of points $\{x_1,\ldots,x_N\}$ is the normalized derivative given by:

$$m_{[1]}(x_{N+1}|\nu,\{x_1,\ldots,x_N\}) = G^{\Xi|Y}[1|\nu,\{x_1,\ldots,x_N\}] = \frac{\frac{\partial^{N+1} G^{\Xi|Y}}{\partial x_1 \ldots \partial x_{N+1}}[1|\nu]}{\frac{\partial^N G^{\Xi|Y}}{\partial x_1 \ldots \partial x_N}[1|\nu]} \tag{36}$$

Writing for the PGFL of the Bayes posterior process with $\nu = \{z_1,\ldots,z_k\}$ gives:

$$m_{[1]}(x_{N+1}|\nu,\{x_1,\ldots,x_N\}) = \frac{\sum_{n=N+1}^{\infty} n(n-1)\ldots(n-N) \int_{S^{n-N-1}} p^{\Xi|Y}(k,z_1,\ldots,z_k,n,x_1,\ldots,x_{N+1},s_{N+2},\ldots,s_n)ds_{N+2}\ldots ds_n}{\sum_{n=N}^{\infty} n(n-1)\ldots(n-N+1) \int_{S^{n-N}} p^{\Xi|Y}(k,z_1,\ldots,z_k,n,x_1,\ldots,x_N,s_{N+1},\ldots,s_n)ds_{N+1}\ldots ds_n} \tag{37}$$

For the special case $N=1$, the ratio of the reduced Palm intensity function of the Bayes posterior process at state $x_2$, conditioned on the existence of a point at state $x_1$ (with $m_{[1]}(x_1) > 0$), and the intensity function of the Bayes posterior process at state $x_2$ is given by:

$$\frac{m_{[1]}(x_2|\nu,\{x_1\})}{m_{[1]}(x_2|\nu)} = \frac{m_{[2]}(x_1,x_2|\nu)}{m_{[1]}(x_1|\nu)m_{[1]}(x_2|\nu)}$$

$$= \frac{\sum_{n=2}^{\infty} n(n-1) \int_{S^{n-2}} p^{Y\Xi}(k,z_1,\ldots,z_k,n,x_1,x_2,s_3,\ldots,s_n)ds_3\ldots ds_n}{\sum_{n=1}^{\infty} n \int_{S^{n-1}} p^{Y\Xi}(k,z_1,\ldots,z_k,n,x_1,s_2,\ldots,s_n)ds_2\ldots ds_n}$$

$$\times \frac{p^Y(k,z_1,\ldots,z_k)}{\sum_{n=1}^{\infty} n \int_{S^{n-1}} p^{Y\Xi}(k,z_1,\ldots,z_k,n,x_2,s_2,\ldots,s_n)ds_2\ldots ds_n} \tag{38}$$

Even for the PPP model where there exists no correlation between the two distinct states $x_1$ and $x_2$, the ratio (38) shows that in the Bayes posterior point process, the conditional intensity function at $x_2$ may depend on $x_1$ based on the structure of $p^{Y\Xi}$. This pairwise correlation and any higher order dependency involving more than two points is lost when a Poisson approximation is used to close a Bayesian recursion.

The ratio of conditional intensity and the intensity functions for a pair of points as in (38) is known in the point process theory literature as the pair correlation function. The pair correlation function is defined as the ratio of the second factorial moment and the first moment at two distinct states [4]. Pair correlation is a non-negative function. Let $G$ denote the PGFL of a finite point process, the pair correlation function $\rho(x_1,x_2)$ is defined by:

$$\rho(x_1, x_2) \equiv \frac{\frac{\partial^2 G}{\partial x_1 \partial x_2}[1]}{\frac{\partial G}{\partial x_1}[1]\frac{\partial G}{\partial x_2}[1]} = \frac{m_{[2]}(x_1, x_2)}{m_{[1]}(x_1) m_{[1]}(x_2)} \quad (39)$$

For a PPP, since the moments factor as in (23), the pair correlation function is identically equal to one. This is a direct result of its independent scattering property. A different pair correlation value indicates dependent scattering between pairs of points. For such a case, if the pair correlation is less than one, the likelihood of observing the pair together in the outcome space is diminished compared to an independent sample model. On the other hand, a pair correlation greater than one indicates a comparatively increased likelihood of observing such a pair together. Therefore, the pair correlation function is an indicator of pairwise interactions in a point process model.

## V. Example: Correlation in the PHD Filter

Let $f^\Xi(s)$ and $\lambda(y)$ be the intensity functions of the predicted target process and the measurement process, both of which are assumed to be Poisson. Let $P^D(s)$ be the target detection probability, and $p(y|s)$ be the measurement likelihood function. The PGFL of the joint event for the PHD filter is derived in [1]. The joint PGFL can be written in the form (see [26]):

$$G^{Y\Xi}[g,h] = \exp\begin{bmatrix} \int_Y \lambda(y)(g(y)-1)dy - \int_S f^\Xi(s)ds \\ + \int_S h(s)(1-P^D(s))f^\Xi(s)ds \\ + \int_S \int_Y g(y)h(s)P^D(s)p(y|s)f^\Xi(s)dy\,ds \end{bmatrix} \quad (40)$$

### A. The Bayes posterior process

Let $\nu = (k, z_1, \ldots, z_k)$ denote the $k$ measurements reported by the sensor. Then the PGFL of the Bayes posterior process $G^{\Xi|Y}$ can be found by carrying out the functional derivatives in (32) on the bivariate PGFL (40). The result is:

$$\begin{aligned} G^{\Xi|Y}[h|\nu] &= \frac{\frac{\partial^k G^{Y\Xi}}{\partial z_1 \ldots \partial z_k}[0,h]}{\frac{\partial^k G^{Y\Xi}}{\partial z_1 \ldots \partial z_k}[0,1]} \\ &= \frac{G^{Y\Xi}[0,h]\prod_{i=1}^k \left[\lambda(z_i) + \int_S h(s)P^D(s)p(z_i|s)f^\Xi(s)ds\right]}{G^{Y\Xi}[0,1]\prod_{i=1}^k \left[\lambda(z_i) + \int_S P^D(s)p(z_i|s)f^\Xi(s)ds\right]} \\ &= \frac{G^{Y\Xi}[0,h]}{G^{Y\Xi}[0,1]} \prod_{i=1}^k \left[\frac{\lambda(z_i) + \int_S h(s)P^D(s)p(z_i|s)f^\Xi(s)ds}{\lambda(z_i) + \int_S P^D(s)p(z_i|s)f^\Xi(s)ds}\right] \end{aligned} \quad (41)$$

where, from (40),

$$\frac{G^{Y\Xi}[0,h]}{G^{Y\Xi}[0,1]} = \exp\left[\int_S (h(s)-1)(1-P^D(s))f^\Xi(s)ds\right] \quad (42)$$

is seen to be the PGFL of the thinned PPP modeling the multitarget process.

The resulting PGFL (41) has a factorized form involving $k+1$ terms. This means that it is the superposition of $k+1$ independent point processes. The first independent target process in (41) is the PPP thinned due to the detection probability which corresponds to the missed targets. Its PGFL is given in (42). The next $k$ independent Bernoulli processes in (41) correspond to the measurements. A Bernoulli process comprises a Bernoulli trial with outcomes labeled "detection" and "missed detection." A target at $x$ is detected with probability $P^D(x)$, and – if detected – it produces the measurement $z_i$ with the likelihood function $p(z_i | x)$. The PGFL of the Bernoulli process for the measurement $z_i$ is

$$G^{\Xi_{Bernoulli}}[h] = \frac{\lambda(z_i) + \int_S h(s) P^D(s) p(z_i | s) f^\Xi(s) ds}{\lambda(z_i) + \int_S P^D(s) p(z_i | s) f^\Xi(s) ds} \tag{43}$$

As a check, note that $G^{\Xi_{Bernoulli}}[1] = 1$.

*B. The pair correlation function*

The pair correlation function of the PHD filter's Bayes posterior process is the ratio of its second factorial moment and the product of its first moments at two distinct target states. In (41), let

$$G^{\Xi|Y}[h|v] = A[h]B[h] \quad \text{where} \quad A[h] = \frac{G^{Y\Xi}[0,h]}{G^{Y\Xi}[0,1]} \tag{44}$$

and

$$B[h] = \prod_{i=1}^{k} \frac{\lambda(z_i) + \int_S h(s) P^D(s) p(z_i | s) f^\Xi(s) ds}{\lambda(z_i) + \int_S P^D(s) p(z_i | s) f^\Xi(s) ds} \tag{45}$$

By using the chain rule, the first derivative of the conditional process can be shown to be the functional:

$$\frac{\partial G^{\Xi|Y}}{\partial x_1}[h|v] = \frac{\partial A}{\partial x_1}[h]B[h] + \frac{\partial B}{\partial x_1}[h]A[h] \tag{46}$$

where

$$\frac{\partial A}{\partial x_1}[h] = A[h](1 - P^D(x_1)) f^\Xi(x_1) \tag{47}$$

$$\frac{\partial B}{\partial x_1}[h] = B[h] \sum_{i=1}^{k} \frac{P^D(x_1) p(z_i | x_1) f^\Xi(x_1)}{\lambda(z_i) + \int_S h(s) P^D(s) p(z_i | s) f^\Xi(s) ds} \tag{48}$$

This can be written as

$$\frac{\partial G^{\Xi|Y}}{\partial x_1}[h|v] = G^{\Xi|Y}[h|v] C_{x_1}[h] \tag{49}$$

where the functional $C_{x_1}[h]$ is:

$$C_{x_1}[h] = (1 - P^D(x_1)) f^\Xi(x_1) + \sum_{i=1}^{k} \frac{P^D(x_1) p(z_i | x_1) f^\Xi(x_1)}{\lambda(z_i) + \int_S h(s) P^D(s) p(z_i | s) f^\Xi(s) ds} \tag{50}$$

Using (10), the first moment of the posterior process can be shown to be:

$$m_{[1]}(x_1 | v) = C_{x_1}[1] = f^{\Xi|Y}(x_1) = (1 - P^D(x_1)) f^\Xi(x_1) + \sum_{i=1}^{k} \frac{P^D(x_1) p(z_i | x_1) f^\Xi(x_1)}{\lambda(z_i) + \int_S P^D(s) p(z_i | s) f^\Xi(s) ds} \tag{51}$$

Similarly, to derive the second factorial moment of the Bayes posterior, the second derivative of the PGFL is needed. Using the chain rule gives the second derivative as

$$\frac{\partial^2 G^{\Xi|Y}}{\partial x_1 \partial x_2}[h|v] = G^{\Xi|Y}[h|v] \left( C_{x_1}[h] C_{x_2}[h] + C_{x_1,x_2}[h] \right) \tag{52}$$

where

$$C_{x_1,x_2}[h] = -\sum_{i=1}^{k} \frac{P^D(x_1) p(z_i | x_1) f^{\Xi}(x_1) P^D(x_2) p(z_i | x_2) f^{\Xi}(x_2)}{\left(\lambda(z_i) + \int_S h(s) P^D(s) p(z_i | s) f^{\Xi}(s) ds\right)^2} \tag{53}$$

An important fact about $C_{x_1,x_2}[h]$ is that it is non-positive for all $x_1$ and $x_2$ for all $h > 0$.

From (52), it is seen that the second factorial moment of the posterior process is equal to:

$$m_{[2]}(x_1, x_2 | \nu) = C_{x_1}[1] C_{x_2}[1] + C_{x_1,x_2}[1] \tag{54}$$

From (51) and (54), the pair correlation function of the Bayes posterior process is

$$\rho(x_1, x_2) = \frac{m_{[2]}(x_1, x_2 | \nu)}{m_{[1]}(x_1 | \nu) m_{[1]}(x_2 | \nu)} = 1 + \frac{C_{x_1,x_2}[1]}{C_{x_1}[1] C_{x_2}[1]} \tag{55}$$

From (55), it is evident that the Bayes posterior process is inherently correlated. Moreover, (53) shows that the interaction term $C_{x_1,x_2}[1]$ is never positive, so $\rho \leq 1$ for any target pair. A process such that $\rho \leq 1$ for all $x_1$ and $x_2$ shows weakly repulsive behavior. This theoretical contribution, first reported in [3], is new to the tracking literature.

Further insights on the nature of pair interactions can be gained by observing (53) and (54). From (54), the second factorial moment of the Bayes posterior process at two distinct target states is equal to the product of the first moments corrected by an interaction term. Writing the interaction term explicitly in (53) shows that this term corrects the double counting of information. The product of the first moment at two distinct target states, i.e. $C_{x_1}[1] C_{x_2}[1]$, includes terms which correspond to the case in which the same measurement is generated by both targets. Intuitively, in the second factorial moment, the job of the interaction term $C_{x_1,x_2}[1]$ is to enforce the single target to single measurement association assumption by subtracting incompatible cases from the product of the first moments. That is, target states in the outcomes of the Bayes posterior process are connected insofar as there is at least one measurement for which they have non-zero likelihoods and detection probabilities. Practically speaking, the interaction between a pair of target states goes to zero as targets become progressively better resolved. Given a PPP prior, pair correlations between distant regions of the target state space are practically negligible. This is a direct result stemming from the PPP target model of the PHD filter.

## VI. Track Extraction Via Reduced Palm Intensity

A Bayesian estimate of $\Xi$ is determined using a specified loss function $L(\zeta | \xi)$. This function gives the associated loss with choosing the estimate $\zeta \in \mathcal{E}(S)$ when the true realization is $\xi \in \mathcal{E}(S)$. The Bayes loss of $\zeta$ is the expected loss, $R(\zeta) = \mathrm{E}[L(\zeta | \xi)]$, where the expectation is with respect to the density $p^{\Xi|Y}(\xi | \nu)$ defined in (33). The Bayes estimate $\hat{\xi}_{Bayes} \in \mathcal{E}(S)$ minimizes the Bayes loss:

$$\hat{\xi}_{Bayes} = \arg\min_{\zeta \in \mathcal{E}(S)} R(\zeta) \tag{56}$$

The Bayes estimate depends on the choice of the loss function $L(\zeta | \xi)$.

In many problems, $L(\zeta | \xi)$ can be specified so that the Bayes estimate reduces to the MAP estimate, $\arg\max_{\xi} p^{\Xi|Y}(\xi | \nu)$. However, the MAP estimate is undefined for the posterior pdf $p^{\Xi|Y}(\xi | \nu)$. This is because events involving different numbers of targets have different units. On the other hand, a two-stage pseudo-MAP estimate can be implemented by using the posterior distribution of the canonical number as well as the posterior intensity functions. In such an approach, the first stage estimates the canonical number and the second stage extracts the estimated number of target states. The statistical consistency of two-stage estimates must be verified on a case by case basis.

Assuming that the pseudo-MAP estimate is reliable, a two-stage sequential track extractor from the Bayes posterior process will be derived. The consistency and efficiency of its estimates should be verified for underlying point process models.

*A. The canonical number estimate*

From (9) the PGF of the canonical number of targets in the Bayes posterior process is given by:

$$F(x) = G^{\Xi|Y}[h|\nu]\big|_{h(\cdot)=x} = \sum_{n=0}^{\infty} x^n p_N^{\Xi|Y}(n|\nu) \tag{57}$$

The pseudo-MAP estimate of the canonical number is then given by:

$$N^{MAP} = \arg\max_n \left( \frac{1}{n!} \frac{\partial^n F[x]}{\partial x^n} \bigg|_{x=0} \right) \tag{58}$$

An alternative (but not theoretically equivalent) estimate of the canonical number is the integral of Bayes posterior intensity function, i.e., the expected number of targets,

$$N^{Expected} = \left[ \int_S f^{\Xi|Y}(s|\nu) ds \right] \tag{59}$$

where, in (59), the brackets $[.]$ is the rounding function which rounds its nonnegative argument to the nearest integer. The estimate (59) is generally simpler to calculate in filter implementations.

*B. Track extraction via reduced Palm intensity*

Conditioned on the estimated canonical number $N$, the pdf of the event $\{x_1,\ldots,x_N\}$ is given by:

$$p^{\Xi|Y}(\{x_1,\ldots,x_N\}|\nu) = N! p^{\Xi|Y}(N,x_1,\ldots,x_N|\nu) = \frac{\partial^N G^{\Xi|Y}}{\partial x_1 \ldots \partial x_N}[0|\nu] \tag{60}$$

The pseudo-MAP track extractor is defined by

$$\xi = \arg\max_{\{x_1,\ldots,x_N\}} \left( p^{\Xi|Y}(\{x_1,\ldots,x_N\}|\nu) \right) \tag{61}$$

The optimization in (61) is over all $N$ target states. The algorithm in Table I gives an approximate pseudo-MAP estimate of the target states corresponding to $N$ peaks for $N \geq 1$. Let $\xi$ be the set of extracted target states, and let $\nu = (k, z_1, \ldots, z_k)$ denote the $k$ measurements reported by the sensor. The algorithm starts with $\xi = \emptyset$ and fills $\xi$ sequentially and recursively with the peaks of the reduced Palm intensity function conditioned on the previously extracted target states. (The iterative procedure of Table I is similar to the SAGE [31] and RELAX [32] algorithms proposed for multiple target detection from the complex radar signal with unthresholded data.)

After the peaks are determined, marginal target pdfs will be calculated. Let $\xi - x_i$ denote the set of extracted target states excluding the target at the state $x_i$, that is, $\xi - x_i = \{x_1,\ldots,x_{i-1},x_{i+1},\ldots x_N\}$. From (8), the marginal pdf of the target in state $x_i$ is given by:

$$p_{x_i}^{\Xi|Y}[x|\nu,\xi-x_i] = \frac{\partial G^{\Xi|Y}}{\partial x}[0|\nu,\xi-x_i] \tag{62}$$

TABLE I
PSEUDO CODE FOR PEAK EXTRACTION ALGORITHM

Given : 1) The set of measurements, $\nu = (k, \{z_1, \ldots, z_k\})$
2) The PGFL of the predicted target process, $G^{\Xi}[h]$

Evaluate PGFL of the Bayes posterior process, $\Xi|Y$

$$G^{\Xi|Y}[h|\nu] = \frac{\partial^k G^{Y\Xi}}{\partial z_1 \ldots \partial z_k}[0,h] \bigg/ \frac{\partial^k G^{Y\Xi}}{\partial z_1 \ldots \partial z_k}[0,1] \qquad \text{Eqn. (33)}$$

Evaluate intensity (PHD, first moment) of $\Xi|Y$

$$f_1^{\Xi|Y}[x|\nu] = \frac{\partial G^{\Xi|Y}}{\partial x}[1|\nu] \qquad \text{(For PHD filter, see Eqn. (50).)}$$

Extract first target:

$$x_1 = \operatorname*{argmax}_{x} f_1^{\Xi|Y}[x|\nu]$$

Let $\xi = \{x_1\}$

Extract targets $2:N$ sequentially from the conditional Palm intensity functions:

For $i = 2 : N$
{

$$f_i^{\Xi|Y}[x_i|\nu, \xi] = \frac{\frac{\partial^i G^{\Xi|Y}}{\partial x_1 \ldots \partial x_i}[1|\nu]}{\frac{\partial^{i-1} G^{\Xi|Y}}{\partial x_1 \ldots \partial x_{i-1}}[1|\nu]} \qquad \begin{array}{l}\text{Eqn. (35)}\\ \text{(For PHD filter, see Eqn. (82)-(84).)}\end{array}$$

$$x_i = \operatorname*{argmax}_{x_i} f_i^{\Xi|Y}[x_i|\nu, \xi]$$

$$\xi = \xi \cup x_i$$

}

VII. TRACK EXTRACTION VIA REDUCED PALM INTENSITY FOR THE PHD FILTER

This section formulates the two-stage sequential track extractor for the PHD filter. Assuming that the predicted target process is a PPP, the MAP estimate for the canonical number is given in Subsection A. Subsection B formulates the sequential peak extraction algorithm of Table I. Subsection C gives the conditional pdf approximation for each extracted target, and Subsection D discusses the computational requirements.

A. *The canonical number estimate*

The PGF of the canonical number of the Bayes posterior process is, using (57) and (41),

$$F(x) = \exp\left[(x-1)\mu_{Missed}\right] \prod_{i=1}^{k} \frac{\lambda(z_i) + x\mu_{Detected}(z_i)}{\lambda(z_i) + \mu_{Detected}(z_i)} \tag{63}$$

where

$$\mu_{Missed} = \int_S (1 - P^D(s)) f^{\Xi}(s) ds \tag{64}$$

and

$$\mu_{Detected}(z_i) = \int_S P^D(s) p(z_i|s) f^{\Xi}(s) ds \tag{65}$$

The derivatives of $F(x)$ evaluated at $x = 0$ give the canonical number pmf. Direct calculation gives

$$p_N^{\Xi|Y}(n|v) = F(0) \sum_{i=0}^{n} \frac{(\mu_{Missed})^{n-i}}{(n-i)!} \sigma_i \left( \frac{\mu_{Detected}(z_1)}{\lambda(z_1)}, \ldots, \frac{\mu_{Detected}(z_k)}{\lambda(z_k)} \right) \tag{66}$$

where $\sigma_i$ denotes the elementary symmetric polynomial of degree $i$. The elementary symmetric polynomial of degree $i$ is formed by adding all distinct products of $i$ distinct terms among its arguments. This leads, from (57), to the MAP estimate for the canonical number:

$$N^{MAP} = \arg\max_{n} p_N^{\Xi|Y}(n|v) \tag{67}$$

Alternatively, the expected number of targets can be estimated from the integral of the intensity function.

*B. Sequential peak extraction*

1) *The first extracted track*

The first target state extracted is defined as the global peak of the posterior intensity function, that is, $x_1 = \arg\max_{x} f^{\Xi|Y}[x|v]$. Subsequent track estimates are conditioned on the first track extracted.

2) *The second extracted track*

Let $\xi = \{x_1\}$. Using (16) the PGFL of the Bayes posterior process conditioned on $\xi = \{x_1\}$ is

$$G^{\Xi|Y}[h|v,\xi] = \frac{\frac{\partial G^{\Xi|Y}}{\partial x_1}[h|v]}{\frac{\partial G^{\Xi|Y}}{\partial x_1}[1|v]} \tag{68}$$

From the conditional PGFL, the reduced Palm intensity is obtained from (17):

$$f^{\Xi|Y}[x|v,\xi] = \lim_{h \to 1} \frac{\frac{\partial^2 G^{\Xi|Y}}{\partial x_1 \partial x}[h|v]}{\frac{\partial G^{\Xi|Y}}{\partial x_1}[1|v]} \tag{69}$$

From (54) and (55), it follows that

$$f^{\Xi|Y}[x|v,\xi] = C_x[1] + \frac{C_{x_1 x}[1]}{C_{x_1}[1]} = \rho(x,x_1)C_x[1] \tag{70}$$

The function $f^{\Xi|Y}[x|v,\xi]$ is the intensity function of the Bayes posterior process reduced by the first extracted target. Direct calculation of (70) gives

$$f^{\Xi|Y}[x|v,\xi] = (1-P^D(x))f^{\Xi}(x) + \sum_{i=1}^{k} \alpha_i \frac{P^D(x)p(z_i|x)f^{\Xi}(x)}{\mu_{z_i}} \tag{71}$$

where

$$\mu_{z_i} = \lambda(z_i) + \int_S P^D(s)p(z_i|s)f^{\Xi}(s)ds \tag{72}$$

and

$$\alpha_i = 1 - \frac{P^D(x_1)p(z_i|x_1)f^{\Xi}(x_1)}{\mu_{z_i} f^{\Xi|Y}[x_1|v]} \tag{73}$$

The structure of (71) can intuitively be seen from the Bayes posterior process of the PHD filter. From (41) it is seen that the posterior process is the superposition of $k+1$ mutually independent processes, namely a PPP for the undetected target process and $k$ Bernoulli processes, one for each measurement. Since the undetected target process is a PPP, the conditional intensity function is unchanged except for being modulated (multiplied) by the probability of missed detection. Conditioned on a target existing at state $x_1$, the intensities at $x$ of the independent Bernoulli processes due to a single measurement $z_i$ are, using (43), seen to be $P^D(x)p(z_i|x)f^{\Xi}(x)/\mu_{z_i}$. These intensities are modulated by the probabilities $\alpha_i$ defined by (73). Equation (71) thus shows that the reduced Palm intensity is the sum of these modulated intensities.

The form of the modulation coefficients (73) is that of a whitening filter. Ideally, the conditional intensity function will be reduced to the background clutter level, i.e., it is a notched spatial filter that reduces the conditional intensity function to the background clutter level in the vicinity of a target that produces a measurement. This theoretical result is new to the tracking literature. Modulation reduces but does not annihilate a peak. In practice, however, the reduced intensity may still have a residual peak around the extracted track. If it does, then this peak may bias the conditional intensities of the other targets.

The second extracted target $x_2$ is

$$x_2 = \arg\max_x f^{\Xi|Y}[x|\nu,\xi=\{x_1\}] \tag{74}$$

In words, the second track is the peak of the reduced Palm intensity function *reduced* by the first extracted track.

3) *The third extracted target track*

To extract the third target, start with the extracted targets $\xi = \{x_1, x_2\}$. The reduced Palm intensity is, from (17),

$$f^{\Xi|Y}[x_3|\nu,\xi] = \lim_{h \to 1} \frac{\frac{\partial^3 G^{\Xi|Y}}{\partial x_1 \partial x_2 \partial x_3}[h|\nu]}{\frac{\partial^2 G^{\Xi|Y}}{\partial x_1 \partial x_2}[1|\nu]} \tag{75}$$

To find the numerator of (75), differentiating (52) leads to

$$\begin{aligned}\frac{\partial^3 G^{\Xi|Y}}{\partial x_1 \partial x_2 \partial x_3}[h|\nu] &= \frac{\partial}{\partial x_3}\left(G^{\Xi|Y}[h|\nu]\left[C_{x_1}[h]C_{x_2}[h]+C_{x_1,x_2}[h]\right]\right) \\ &= G^{\Xi|Y}[h|\nu]\begin{pmatrix}C_{x_1}[h]C_{x_2}[h]C_{x_3}[h]+C_{x_1,x_2}[h]C_{x_3}[h]\\+C_{x_1,x_3}[h]C_{x_2}[h]+C_{x_2,x_3}[h]C_{x_1}[h]+C_{x_1,x_2,x_3}[h]\end{pmatrix}\end{aligned} \tag{76}$$

where the previously undefined term $C_{x_1,x_2,x_3}[h]$ in (76) is given by

$$\begin{aligned}C_{x_1,x_2,x_3}[h] &= \frac{\partial C_{x_1,x_2}[h]}{\partial x_3} \\ &= \sum_{i=1}^{k} \frac{2\prod_{j=1}^{3}P^D(x_j)p(z_i|x_j)f^{\Xi}(x_j)}{\left(\lambda(z_i)+\int_S h(s)P^D(s)p(z_i|s)f^{\Xi}(s)ds\right)^3}\end{aligned} \tag{77}$$

Using (10), the conditional intensity at $x_3$ is

$$f^{\Xi|Y}[x_3|\nu,\xi] = \left.\frac{\frac{\partial^3 G^{\Xi|Y}}{\partial x_1 \partial x_2 \partial x_3}[h|\nu]}{\frac{\partial^2 G^{\Xi|Y}}{\partial x_1 \partial x_2}[1|\nu]}\right|_{h(.)=1} = \frac{A}{B} \tag{78}$$

where

$$A = C_{x_1}[1]C_{x_2}[1]C_{x_3}[1] + C_{x_1,x_2}[1]C_{x_3}[1] + C_{x_1,x_3}[1]C_{x_2}[1] + C_{x_2,x_3}[1]C_{x_1}[1] + C_{x_1,x_2,x_3}[1]$$
$$B = C_{x_1}[1]C_{x_2}[1] + C_{x_1,x_2}[1]$$
(79)

Therefore the exact reduced Palm intensity function for the target pair is

$$f^{\Xi|Y}[x_3 | \nu, \xi] = C_{x_3}[1] + \frac{C_{x_1,x_3}[1]C_{x_2}[1] + C_{x_2,x_3}[1]C_{x_1}[1] + C_{x_1,x_2,x_3}[1]}{C_{x_1}[1]C_{x_2}[1] + C_{x_1,x_2}[1]}$$
(80)

Comparing (80) to (70), it can be seen that the computation required to find the subtraction term for the intensity function for two targets increases significantly compared to that of a single target. This is because three terms are needed to calculate the reduction of intensity due to a measurement $z_i$ since it could be generated by $x_1$ and not $x_2$, by $x_2$ and not $x_1$, or by neither.

4) *Higher order extracted tracks*

The computational complexity of the reduced Palm intensity function with events including higher number of extracted targets such as $\xi = \{x_1, \ldots, x_l\}$ grows too rapidly for practical implementation. Explicit expressions can be found, but they are not given here.

A useful approximation assumes that the measurement system behaves so that for a detected target at $x$, the likelihood function of its measurement $z_i$ is much greater than the likelihood of any other measurement $z_j$ in the vicinity of $x$, that is, $p(z_i | x) \gg p(z_j | x)$. The factorial moments of the Bayes posterior process factor approximately around the detected target locations:

$$m_{[l]}(x_1, \ldots, x_l) \approx \prod_{i=1}^{l} C_{x_i}[1]$$
(81)

The approximation (81) assumes, in effect, that regions of high measurement likelihood do not intersect. That is, (81) means that, in practice, the support of the intensity function can be partitioned into measurement-specific, non-overlapping regions. Issues due to finite sensor resolution for close targets are ignored. Accordingly, the sequential application of the single target reduction for all extracted targets $\xi = \{x_1, \ldots, x_N\}$ gives the following approximation to the reduced Palm intensity function in the vicinity of detected targets:

$$f^{\Xi|Y}[x|\nu,\xi] \approx C_x[1] + \sum_{j=1}^{N} \frac{C_{x_j,x}[1]}{C_{x_j}[1]}$$
(82)

The negative contribution $C_{x_j,x}[1]/C_{x_j}[1]$ due to each extracted target $x_j$ in (82) will be referred as the Palm corrector for a target at $x_j$. Writing (82) explicitly gives[2]:

$$f^{\Xi|Y}[x|\nu,\xi] \approx (1-P^D(x))f^{\Xi}(x) + \sum_{j=1}^{k} \frac{P^D(x)p(z_j|x)f^{\Xi}(x)}{\mu_{z_j}}$$
$$- \sum_{l=1}^{N}\sum_{j=1}^{k} \frac{\frac{P^D(x)p(z_j|x)f^{\Xi}(x)P^D(x_l)p(z_j|x_l)f^{\Xi}(x_l)}{\mu_{z_j}^2}}{(1-P^D(x_l))f^{\Xi}(x_l) + \sum_{i=1}^{k}\frac{P^D(x_l)p(z_i|x_l)f^{\Xi}(x_l)}{\mu_{z_i}}}$$
(83)

where $\mu_{z_j}$ is defined in (72). Similar to (71), (83) can be rewritten as

$$f^{\Xi|Y}[x|\nu,\xi] \approx (1-P^D(x))f^{\Xi}(x) + \sum_{j=1}^{k} \alpha_j \frac{P^D(x)p(z_j|x)f^{\Xi}(x)}{\mu_{z_j}} \tag{84}$$

where $\alpha_j$ is the Palm modulation term given by (85).

---

[2] The first order Palm corrector approximation may lead to numerical issues (e.g. negatively weighted particles in a sequential Monte Carlo implementation) when (80) does not hold. In practice, this problem can be avoided by replacing negative weights by zero.

$$\alpha_j = 1 - \sum_{l=1}^{N} \frac{P^D(x_l)p(z_j|x_l)f^{\Xi}(x_l)}{\mu_{z_j} f^{\Xi|Y}[x_l|\nu]} \tag{85}$$

### C. Conditional distributions for extracted tracks

After the peaks are determined, the final task of the track extractor is the estimation of a conditional target pdf for each extracted target. From (52) and using the approximation (81), the pdf of the target $x_i$ conditioned on the other extracted targets, $\xi - x_i = \{x_1,...,x_{i-1},x_{i+1},...x_N\}$, is given up to a proportionality constant $K$ by

$$p_{x_i}^{\Xi|Y}[x|\nu,\xi-x_i] \approx \frac{1}{K}\left( C_x[0] + \sum_{\substack{j=1 \\ j \neq i}}^{N} \frac{C_{x_j x}[0]}{C_{x_j}[0]} \right) \tag{86}$$

In comparison to the conditional intensity function of (82), the evaluation of (86) is done by taking $h(.) = 0$ [3]. This difference between the conditional intensity function and the conditional target pdf (up to a proportionality constant) can be shown using (8) and (16). Starting with the Bayes posterior target process, defined in (34), the reduced Palm process, defined in (16), gives the reduced process conditioned on other extracted targets. Taking the next functional derivative and setting $h(.)=0$ as in (8) annuls the contribution of events with more than $N$ targets to the estimated conditional target pdf. On the other hand, taking the next functional derivative and setting $h(.) \to 1$ would allow the contribution of events with more than $N$ targets to the conditional intensity function. Writing (86) explicitly gives:

$$p_{x_i}^{\Xi|Y}[x|\nu,\xi-x_i] \approx \frac{1}{K}\left[ \begin{array}{c} (1-P^D(x))f^{\Xi}(x) + \sum_{j=1}^{k} \frac{P^D(x)p(z_j|x)f^{\Xi}(x)}{\lambda(z_j)} \\ -\sum_{\substack{l=1 \\ l \neq i}}^{N}\sum_{j=1}^{k} \frac{\frac{P^D(x)p(z_j|x)f^{\Xi}(x)P^D(x_l)p(z_j|x_l)f^{\Xi}(x_l)}{\lambda(z_j)^2}}{(1-P^D(x_l))f^{\Xi}(x_l) + \sum_{i=1}^{k}\frac{P^D(x_l)p(z_j|x_l)f^{\Xi}(x_l)}{\lambda(z_j)}} \end{array} \right] \tag{87}$$

Equation (87) can be rewritten in the in the Palm modulation form similar to (84):

$$p_{x_i}^{\Xi|Y}[x|\nu,\xi-x_i] \approx \frac{1}{K}\left[ (1-P^D(x))f^{\Xi}(x) + \sum_{j=1}^{k} \alpha_j \frac{P^D(x)p(z_j|x)f^{\Xi}(x)}{\lambda(z_j)} \right] \tag{88}$$

where

$$\alpha_j = 1 - \sum_{\substack{l=1 \\ l \neq i}}^{N} \frac{P^D(x_l)p(z_j|x_l)f^{\Xi}(x_l)}{\lambda(z_j)\gamma_{z_j}} \tag{89}$$

and

---

[3] That is, $p_{x_i}^{\Xi|Y}[x|\nu,\xi-x_i]$ is proportional to the Papangelou conditional intensity function, see [29] and [5, Eq. 15.5.1a].

$$\gamma_{z_j} = \left(1 - P^D(x_l)\right) f^{\Xi}(x_l) + \sum_{i=1}^{k} \frac{P^D(x_l) p(z_j | x_l) f^{\Xi}(x_l)}{\lambda(z_j)} \quad (90)$$

Both (84) and (89) show that the function of the Palm corrector is to modulate the weights of the Bernoulli process components due to target detections. The general structure of the Bayes posterior process does not change, except for the weights of individual components. For example, in the Gaussian mixture PHD filter implementation of [33], the application of the Palm corrector alters only the weight, not the mean or the covariance of an individual Gaussian component.

It should be noted that the term $f^{\Xi}(x_l) \neq 0$ cancels from both (89) and (90). In a similar fashion, the term $f^{\Xi}(x)$ cancels from both (84) and (85). That is, the Palm corrector's modulation of measurement weights does not depend on the intensity of the conditional point insofar as $f^{\Xi}(x_l) \neq 0$.

In a practical application, the pdf defined by (88) should be truncated around the extracted peak to avoid the biasing effect from distant targets. This was done according to the sensor characteristics in the sequential Monte Carlo implementation of [22]. For the Gaussian mixture PHD (GM-PHD) implementation, the tree structure that holds the track pedigrees recommended in [34] provides a natural grouping of components. Alternatively, for the Gaussian mixture implementation, mixture components that share at least a single common measurement in a history of last $S$ scans may be clustered together. The extracted track pdf will then be truncated to those components that share at least a single common measurement with the peak in the last $S$ scans.

*D. Computational requirements*

Let $N$ be the number of existing targets for a given scan estimated according to (67), or by integrating the intensity function of the Bayes posterior process, the peak extraction algorithm of Table I requires, on average, $N$ sequential optimizations. The first optimization searches for the peak of the intensity function, which is the direct output of the PHD filter, whereas the remaining $N-1$ optimizations search for the peaks of the reduced intensity function, which are calculated by the recursive application of the Palm corrector. The first order Palm corrector for $N$ targets, defined in (83) and rewritten in (84)-(85), is structurally similar to the intensity function of the Bayes posterior process defined in (51), but with an additional modulation coefficient for each measurement. Therefore, the reduction of $N$ peaks from the posterior intensity function, at worst, requires the sequential recalculation of the Bayes posterior intensity function for each extracted peak. Each sequential peak reduction includes an additional modulation coefficient which is calculated for each measurement. In practice, however, only the measurements close to the peaks will have an effect on the results, thus computational savings can be achieved by gating procedures. Moreover, for the GM-PHD implementation, it is only required to calculate the modulation coefficients. That is, the weight, not the mean or the covariance of the Gaussian component itself is altered in recursions.

After the peaks of the Bayes posterior process are determined, the track extraction algorithm will determine the pdfs for extracted tracks. Each pdf due to an extracted peak is calculated by conditioning on the existence of targets at the remaining $N-1$ peaks. The conditional pdf of (87) is quite similar to the conditional intensity function of (84), but with altered weights. Therefore, the extraction of individual target pdfs will require, at worst, a computation similar to carrying out $N^2 - N$ times the PHD filter's measurement update formula.

VIII. NUMERICAL EXAMPLES

For illustration purposes, we consider a scenario involving two targets moving in close proximity. The target state vector $[x \ \dot{x} \ y \ \dot{y}]^T$ is composed of position (m) and velocity components (m/s). Observations are the position components of target states corrupted by additive Gaussian noise. The surveillance region in which observations are generated has a field of view extending $\pm 2000$ m in both position coordinates. Clutter is assumed to be PPP with uniform intensity over the surveillance region. Figure 1 shows the simulated target paths in the scenario. Targets are initiated with state vectors $\begin{bmatrix} -1700 & 0 & 1000 & -50 \end{bmatrix}^T$ for the first target and $\begin{bmatrix} -1700 & 0 & -1000 & 50 \end{bmatrix}^T$ for the second target. In the first 30 seconds of the scenario, targets approach each other slowly by performing a coordinated turn at 3 degrees per second. For the next 30 seconds, they move with constant speed in parallel separated by 90 meters. Finally, they drift apart by carrying out another 90 degree turn in the last 30 seconds.

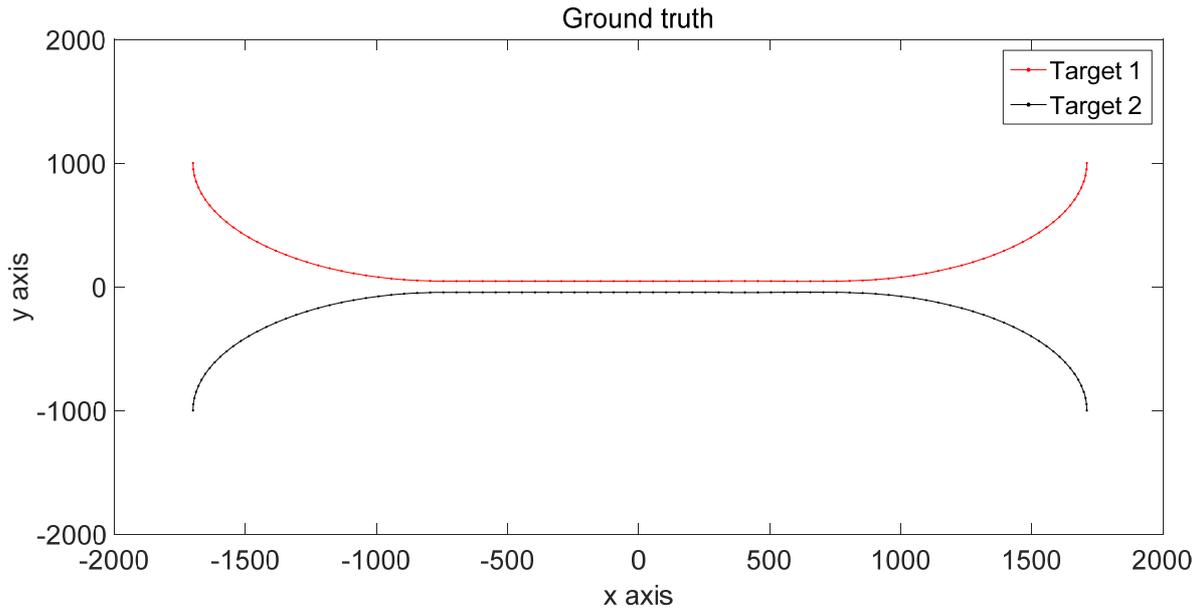

Fig. 1. True target positions.

In the tracking filter, target motion is modeled according to the linear Gaussian dynamics. If $x(k)$ represents the state vector of a target at time instant $k$, then

$$x(k+1) = Fx(k) + \upsilon(k)$$
$$F = \begin{bmatrix} 1 & T & 0 & 0 \\ 0 & 1 & 0 & 0 \\ 0 & 0 & 1 & T \\ 0 & 0 & 0 & 1 \end{bmatrix} \quad (91)$$

where $\upsilon(k)$ is a Gaussian distributed random variable with zero-mean and covariance matrix $Q$,

$$Q = \begin{bmatrix} \frac{T^3}{3} & \frac{T^2}{2} & 0 & 0 \\ \frac{T^2}{2} & T & 0 & 0 \\ 0 & 0 & \frac{T^3}{3} & T \\ 0 & 0 & 0 & T \end{bmatrix} \sigma_P^2 \quad (92)$$

and where the process noise standard deviation $\sigma_P$ of $\upsilon(k)$ is taken to be 5 m/s$^2$.

Measurements from detected targets are assumed to be linear Gaussian, that is, the measurement $z_k$ of a target at $x(k)$ is a realization of the model

$$Z(k) = Hx(k) + w(k)$$
$$H = \begin{bmatrix} 1 & 0 & 0 & 0 \\ 0 & 0 & 1 & 0 \end{bmatrix} \quad (93)$$

where the error term $w$ is zero mean Gaussian random variable with covariance matrix $R = I_2 \sigma_M^2$, where $I_2$ is the identity matrix of size 2 and $\sigma_M$ is the measurement error standard deviation, which is taken to be 25 m. Target detection probability is constant and independent of target state. Issues of sensor resolution are ignored. Measurement update rate is 1 Hz.

The first target tracker is a sequential Monte Carlo implementation of the PHD filter. In this implementation, the target birth and the target spawn conditions are not considered. It is assumed that existing targets survive till the end of the scenario with a probability of one. The initial intensity function is a Gaussian mixture generated by two Kalman filters using the first 10 target originated measurements correctly assigned to each target. The initiation of Kalman filters is made according to the two point differencing method of [35]. In the initiation scans, both targets generate a measurement at all times. At the end of the 10$^{th}$ scan, 20,000 particles are sampled from each Gaussian corresponding to a target. Concatenating 40,000 particles and setting the expected number of targets to 2 give the particle approximation of the initial intensity function of the target process. To be clear, the prior target process is assumed to be a PPP whose intensity function is the superposition of two Gaussian components produced by the Kalman filters. From scan 11 to scan 91, the unweighted particle approximation of the target process' intensity function is propagated using the PHD filter's recursions. That is, each particle is predicted in time according to the target motion model defined by (91)-(92) (i.e. for the time update, transition prior probability distribution is the importance function), and after each measurement update, which is carried out according to (51), weighted particles are resampled to give rise to 40,000 unweighted particles. Additionally, added to each resampled particle is a small additive random dithering sampled from a zero mean Gaussian with a covariance matrix $Q_{Dither}$ defined by

$$Q_{Dither} = \begin{bmatrix} 0.33 & 0.5 & 0 & 0 \\ 0.5 & 1 & 0 & 0 \\ 0 & 0 & 0.33 & 0.5 \\ 0 & 0 & 0.5 & 1 \end{bmatrix} \quad (94)$$

Figure 2, shows the position distribution of Monte Carlo particles taken from a single run for this implementation of the PHD filter. In Figure 2, a random subset of particle cloud is plotted for every 5 scans beginning with scan 11.

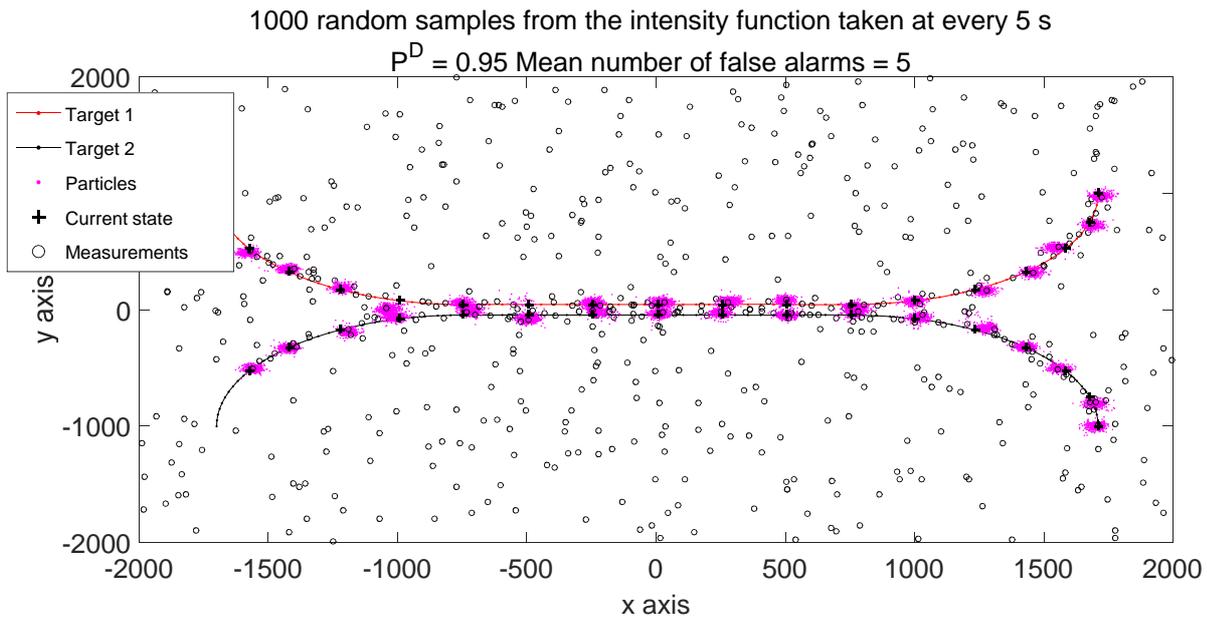

Fig. 2. Sequential Monte Carlo implementation of the PHD filter. Target particles are plotted at every five seconds beginning with the 11$^{th}$ scan. 40,000 particles approximating the intensity function of the target process are down sampled to 1000.

Applied to this filter is the track extracting algorithm presented in Table 1. In the implementation of the track extractor, the canonical number is calculated according to (59). To determine the peaks of the Bayes posterior intensity function, similar to [36], the surveillance region is divided into 51 x 51 resolution cells of equal size; that is 80 m x 80 m for both position coordinates. Each resolution cell is subdivided into 21 x 21 sub-resolution cells with a size of 4 m x 4 m. The total weight of particles falling inside each cell gives the weight for that cell. The global peak of the intensity function of the Bayes posterior process is approximated at the state of particle that has the largest weight and which falls inside the heaviest resolution as well as the sub resolution cells. (Alternatively, the weighted mean of these particles could be used.) While not implemented for this demonstration, a full grid that includes both position and velocity components may improve peak detection, especially for crossing targets.

Figure 3 shows the position distribution of extracted tracks for a single run (red particles correspond to the first extracted track and blue particles correspond to the second extracted track). Observing Figure 3, it can be seen that the location of the global peak of the Bayes posterior intensity function, i.e. the first extracted track, switches between Target 1 and Target 2. Because there are (on average) two peaks of equal height, the global peak oscillates randomly between targets based on the distribution of Monte Carlo particles. Nevertheless, the subtractive (spatial whitening) nature of the reduced Palm intensity function predicted by (70) can be observed from the separation of particles of the extracted tracks. To see this more clearly, Figure 4 zooms in scan 51. Plotted are (a) the intensity function of the Bayes posterior process, (b) superposition of the reduced Palm intensity functions, (c) the reduced Palm intensity function around the global peak, and (d) the reduced Palm intensity function around the secondary peak. Figure 4 verifies that, as theory indicates in the equation (70), the reduced Palm intensity function subtracts intensity due to the conditioned target from the Bayes posterior intensity function. This in effect divides a single particle cloud generated by two targets into two near-by but separate groups. It is interesting (and potentially important for applications) to note that the target-specific particle clouds are overlapped, and not divided into non-overlapping clouds as they would be if a classifier had been used to determine a discriminant boundary.

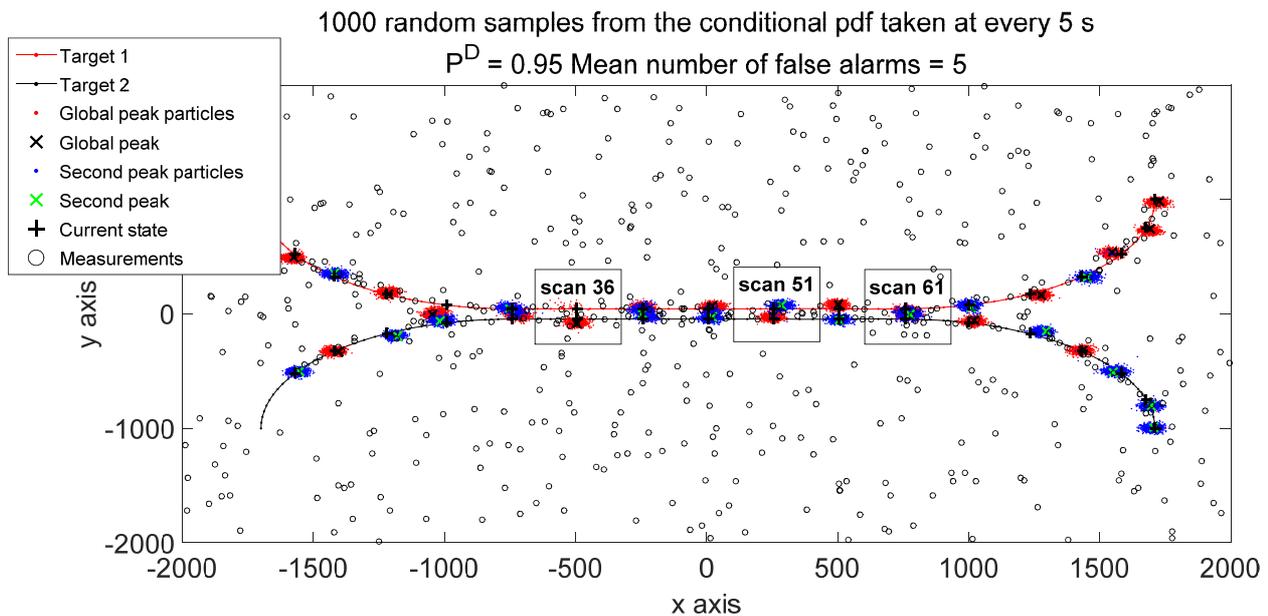

Fig. 3. Sequential Monte Carlo implementation of the PHD filter. Target particles are plotted at every fifth scan beginning with the scan 11. 40,000 particles approximating the conditional probability density function due to each target are down sampled to 1000.

On the other hand, as shown in Figure 5, Palm track separation is unsuccessful when the Bayes posterior intensity function fails to reveal two peaks. Figure 5 zooms in scan 61. Even though there are two target originated measurements in this scan, the reduced Palm intensity function does not separate the tracks. This is because the target measurements are close to each other with respect to measurement error standard deviation. (There is 20 m separation in x axis and 29 m separation in y axis between measurements, and the measurement error standard deviation is 25 m). Conditioned on the global peak, the reduced Palm intensity function has a peak practically at the same location. Since two target states used for conditioning are practically at the same state, the sequential peak extraction algorithm of Table 1 fails to find two distinct peaks. It is important to note, however, that the failure of the Bayes posterior intensity function to reveal two peaks does not cause a catastrophic failure in the Palm track extractor. A natural concern would be that conditioning on one of two nearby targets would cause the reduced Palm intensity to subtract both targets. This does not happen in Fig. 5.

Another case involving undetected target can also be seen in Fig. 3 in the box that corresponds to scan 36. In this scan, upon close inspection, it is seen that Target 2 is detected, but Target 1 is missed. Because of the well-known target death problem for the PHD filter [37], the intensity function is quickly reduced in the vicinity of Target 1. As a result, the canonical number estimator reports a single target, so the peak extractor finds a single peak around Target 2's location. That means that Target 1 is not extracted and the spread of particles for Target 2 is biased slightly upwards towards Target 1 to accommodate for the presence of Target 1 in the prior from the previous scan.

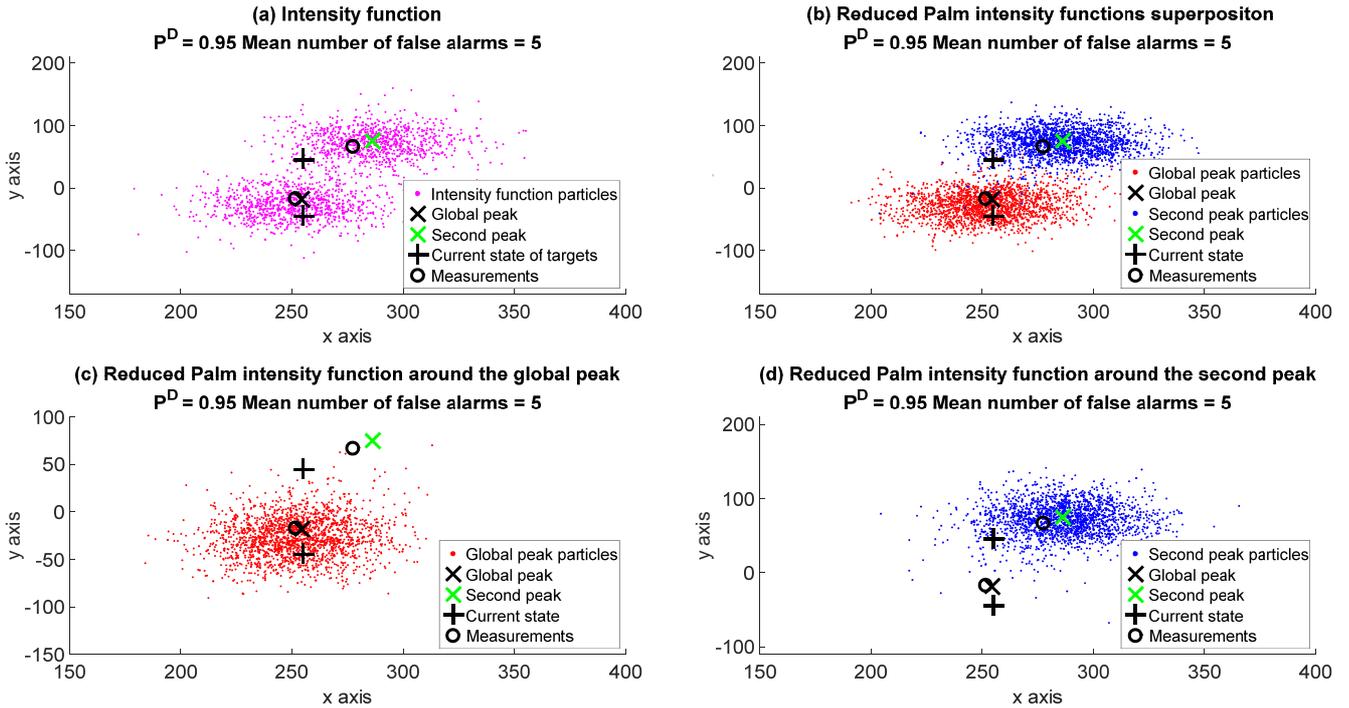

Fig. 4. Scan 51 results. Intensity functions are down sampled to 1000 particles. (a) The Bayes posterior intensity function, (b) Superposition of reduced Palm intensity functions due to two peaks (c) The reduced Palm intensity function around the global peak (d) The reduced Palm intensity function around the second peak

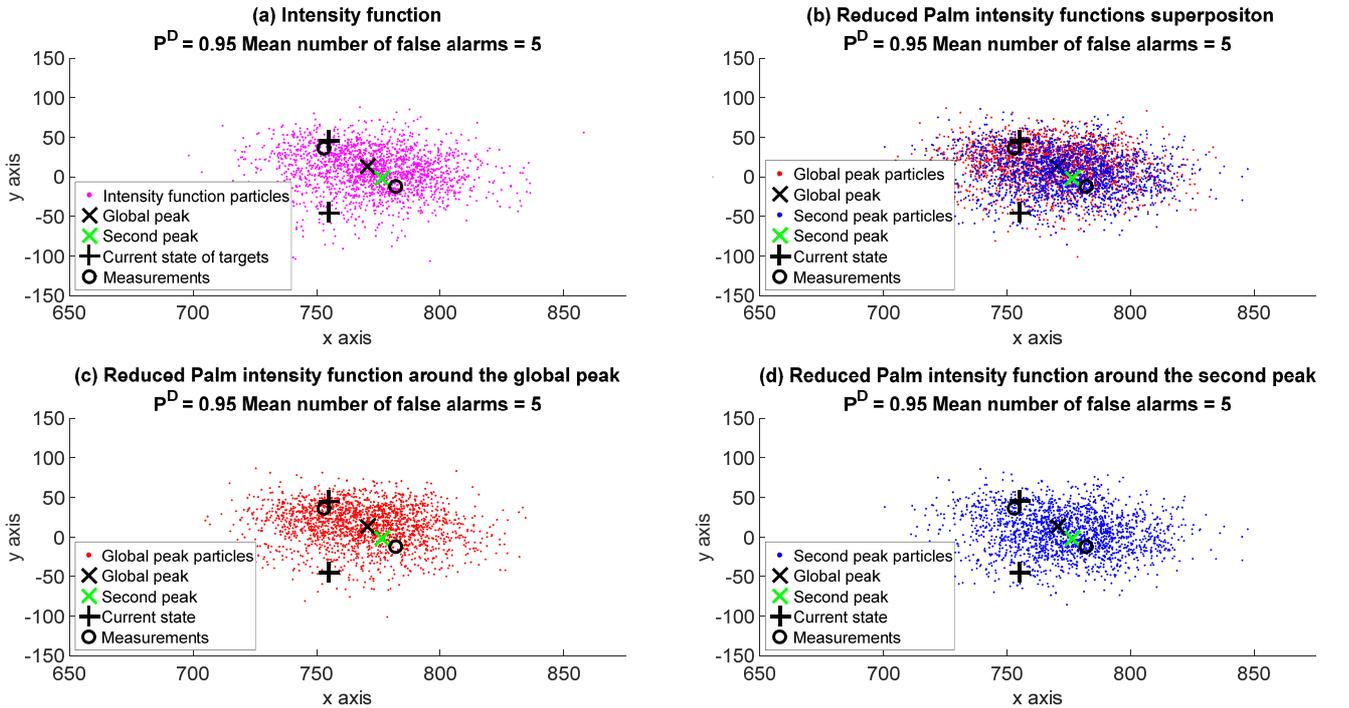

Fig. 5. Scan 61 results. Intensity functions are down sampled to 1000 particles. (a) The Bayes posterior intensity function, (b) Superposition of reduced Palm intensity functions due to two peaks (c) The reduced Palm intensity function around the global peak (d) The reduced Palm intensity function around the second peak

From these examples, it is clear that the performance of the Palm track extractor depends on the performance of both the canonical number estimator and the performance of the peak detector. It is well-known that the canonical number estimate of the PHD filter is unstable [37], [38], [39]. Even disregarding errors due to the clutter or misdetections, inspection of Figure 3 shows that under a benign condition, the global peak of the PHD intensity function jumps between targets. The situation deteriorates rapidly with increasing the clutter rate or decreasing the level of the probability of detection. It is evident that the effective use of the Palm method as a track extractor requires stabilizing the global peak of a target. There are numerous studies concerning stabilization of PHD filter estimates. For example, [34] proposes a tree structure similar to track oriented MHT for a Gaussian mixture implementation of the PHD filter, while [36] and [40] use track labeling and track to measurement assignment for a sequential Monte Carlo implementation. The present study is intended for the introduction of reduced Palm intensity function, so discussion of methods for stabilizing the PHD filter peaks is outside the scope of this paper—it is an active area of research and the subject of on-going work.

To demonstrate the potential capability of the Palm track extractor with numerical results, a comparison with the GM-PHD implementation of [33] is carried out in this scenario. The GM-PHD filter is chosen for comparison because its intensity function approximation is inherently clustered, therefore its track extractor is both more efficient and more robust (see [33] and [34]) in comparison to the clustering methods suggested for the target state estimation in sequential Monte Carlo implementations. As done in the sequential Monte Carlo tracker example, the initial intensity function of the GM-PHD implementation is a Gaussian mixture involving two components generated by two Kalman filters using the first 10 target originated measurements correctly assigned to each target. No further birth or spawn components are generated in the following scans and the probability of target survival is assumed to be one. For the management of Gaussian components in the GM-PHD filter, the prune and merge thresholds are, respectively, $T = 10^{-5}$ and $U = 4$. The maximum number of Gaussian components is set to $J_{max} = 500$ (see Table II in [33] for the meanings of these parameters). Track extraction for the GM-PHD filter is carried out according to Table III of [33], and the minimum component weight for the track extraction is set to 0.5. The Palm track extractor is implemented separately without altering the GM-PHD recursions. On the other hand, the Palm track extractor works before the Bayes posterior process is approximated with merge and prune operations at each measurement update, while the GM-PHD state extractor works on the approximated Bayes posterior intensity function.

The canonical number estimator for the Palm track extractor is the expected target number rounded to the nearest integer according to (59). Following the recommendation in [33], the peak search algorithm of the Table I is altered to work with the weight, not the height, of Gaussian components. To be clear, the first peak extracted is the mean of the heaviest Gaussian component. Conditioned on this peak, reduced Palm process is calculated by modulating the weights of detected Gaussian components according to (84) and (85). The next extracted peak is the heaviest component of the reduced Palm intensity function. This procedure is repeated until all peaks are determined. The state estimate for each extracted target is the expected target state calculated according to the pdf defined by (87). For each extracted Gaussian peak, this pdf is truncated to involve only those components that were updated by at least a single common measurement within a history of 5 scans.

Figure 6 shows 20000 Monte Carlo run average results of the mean optimal subpattern assignment (MOSPA) statistic with parameters $p = 2$ and $c = 200\,\text{m}$ (see [41], [42] for the definition of MOSPA) and the mean number of extracted tracks. To generate Figure 6 from the scenario described previously and shown in Figure 1, in each Monte Carlo run measurements are created independently with the target detection probability level set to 0.98 and the mean number of false alarms set to 10. Three salient features are observable in Figure 6. The first feature is that there is a visible negative bias in the reported number tracks for the GM-PHD state extractor when targets are in close proximity. The negative bias is especially more pronounced when targets are approaching each other, whereas target separation is more graceful. In contrast, the Palm track extractor's canonical number estimates do not show such pronounced bias. The next salient feature is that, while there is no noticeable difference in MOSPA statistic for both state estimators when targets are far apart, the MOSPA statistic for both state estimators show an increase in the error level when targets are at close proximity. The last salient feature is that, although the MOSPA statistic for the Palm track extractor is lower on the average, its level is higher than the GM-PHD state estimator's level while targets are approaching to each other or when they are separating from each other. This suggests that, target state estimates of the Palm track extractor are biased by the Gaussian components belonging to the other target. This problem is in part influenced by the ad-hoc target pdf truncation method which involves clustering components that share at least a single common measurement within a history of 5 scans. It may be ameliorated with more advanced track management methods such as the one recommended in [34].

Figure 7 shows 20000 Monte Carlo run results with target detection probability lowered to 0.9. At this level, both track extractors show a clearly visible negative bias in the estimated number of tracks. While the canonical number estimator for the GM-PHD filter show similar features in comparison to Figure 6, there exist a noticeable difference in the canonical number estimates in comparison to the Palm track extractor even when targets are far apart from each other. The negative bias in the reported number of tracks after the rounding operation can be analytically explained for the case with no false alarms. Assuming that there is a single target which is detected according a known target detection probability, $P^D$, and the prior intensity function integrates to one, it is straightforward to show that if the target produces a measurement the Bayes posterior intensity function will integrate to $2 - P^D$, whereas if the target is missed, the Bayes posterior intensity function will integrate to $1 - P^D$. Taking the expectation shows that the canonical number of the Bayes posterior process is unbiased. On the other hand, rounding operation biases the reported number of extracted tracks. For example, when $P^D > 0.5$, if the target is detected, the expected target number will be rounded to 1, whereas if the target is missed, the expected target number will be rounded to 0. Therefore, the expected number of reported tracks after the rounding operation is equal to $P^D$, which is negatively biased. In contrast, for $P^D < 0.5$, the expected number of reported tracks after the rounding operation is equal to $1 + P^D$, which is positively biased. The GM-PHD implementation puts another stress to the already biased canonical number estimates by rounding the weights of individual Gaussian components. The detrimental effect is especially more pronounced when multiple prior Gaussian components compete for the same measurement.

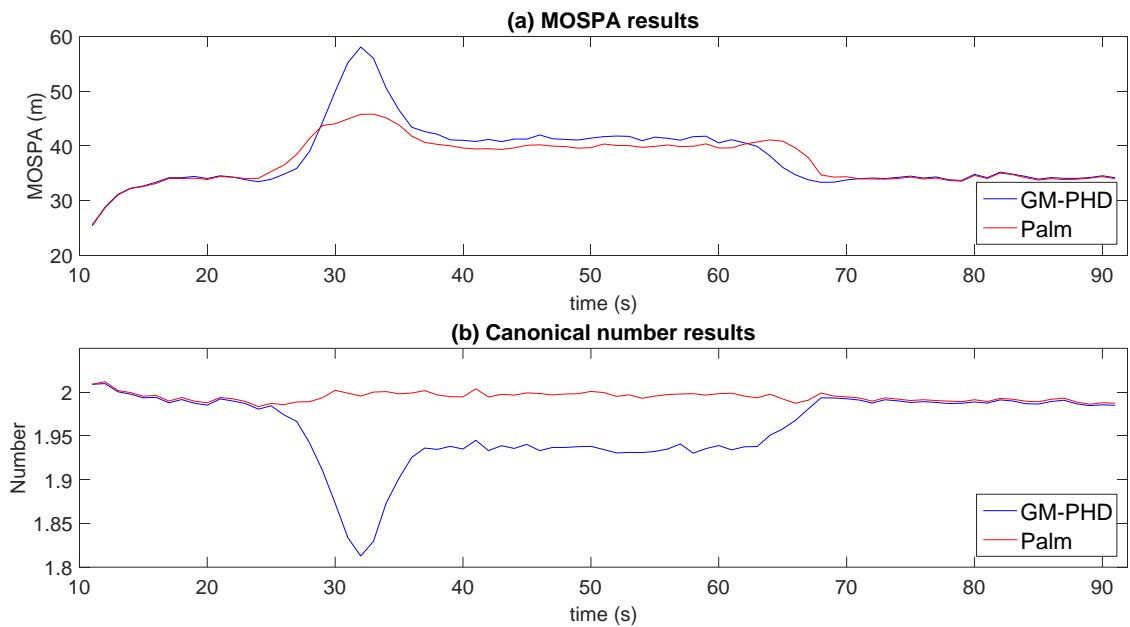

Fig. 6. 20000 Monte Carlo results for the MOSPA and the canonical number estimate with the target detection probability at 0.98 and the mean number of false alarms at 10.

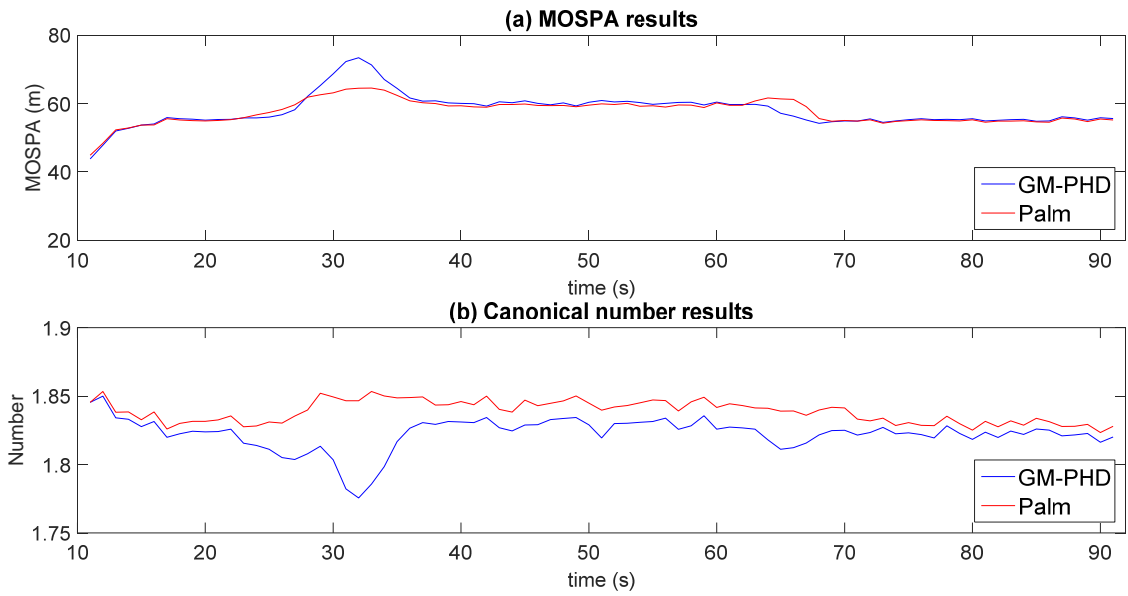

Fig. 7. 20000 Monte Carlo results for the MOSPA and the canonical number estimate with the target detection probability at 0.90 and the mean number of false alarms at 10.

Figure 8 shows the scenario average MOSPA results from 1000 Monte Carlo runs at varying target detection probability and false alarm rates. From Figures 6 and 7, it is seen that the Palm track extractor has uniformly lower MOSPA results in comparison to the GM-PHD filter's state extractor. One reason for this is that the GM-PHD filter's state extractor has a bias on its canonical number estimates due to the way it thresholds individual Gaussian components. (Ways to reduce this bias are not discussed in [33] but were subsequently developed in the context of other trackers.) The examples of this section are a first demonstration that the reduced Palm process is a new and valuable tool for separating individual target states.

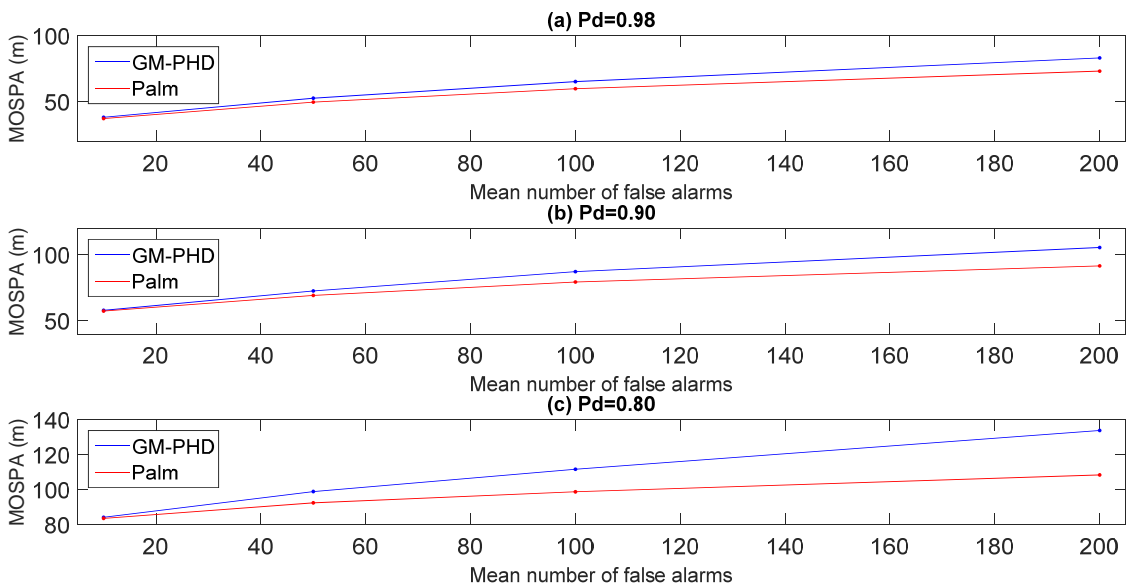

Fig. 8. 1000 Monte Carlo run average MOSPA results at the probability of detection level (a) 0.98, (b) 0.90, (c) 0.8 and mean number of false alarms varying from 10 to 200.

IX. CONCLUSION

This paper investigates pair correlations in the Bayes posterior target process for multitarget tracking filters using finite point process target models. The pair correlation function is the second factorial moment at two distinct target states divided by the product of first moments. This study uses the pair correlation function to show that the PPP target model of the PHD filter leads to a spatially correlated Bayes posterior target process that has weakly repulsive target pair interactions.

A closely related concept to the pair correlation function is the Palm distribution. The Palm distribution defines the conditional distribution of points, given that a point exists in all possible realizations that can be produced by the finite point process model. For target tracking purposes, this means that the Palm process defines the multitarget process given the state of a target that is known to exist. Another related distribution is the reduced Palm distribution. The reduced Palm distribution defines the distribution of points that can be produced by the finite point process model after *excluding* the point used in the conditioning. For target tracking purposes, the reduced Palm distribution defines the mathematically correct way to remove a target from a point process target model. This paper describes the PGFL of the reduced Palm process. With this tool at hand, it then derives a two-stage pseudo-MAP track extractor, namely the Palm track extractor, for a general finite point process target model.

The Palm track extractor is a two-stage algorithm. In the first stage the number of targets is estimated. In the second stage peaks are extracted sequentially by removing a single target which is estimated at the peak of the reduced Palm intensity function conditioned on the previously extracted targets. To solidify the method, a first order approximation of the Palm track extractor which considers only pairwise interactions among targets is formulated for the special case of the PHD filter.

The approximate Palm track extractor for the PHD filter is applied in a simulated numerical study which involves tracking two close-by targets. The experiment is carried out using a sequential Monte Carlo implementation of the PHD filter. It shows that the Palm track extractor can extract separate target pdfs under sparse clutter if the target peaks are resolved in the posterior intensity function. A second experiment is carried out using the same scenario, but with a GM-PHD filter implementation. Numerical results show improvements in track extraction with the use of Palm method. An on-going study investigates Palm filters which propagate the reduced Palm intensity function with extracted tracks. Conditional intensity functions for target tracking filters using multiple Bernoulli and IID clusters target models are also currently being investigated.